\documentclass[prb,twocolumn,showpacs]{revtex4}
\usepackage{graphicx}

\newcommand{\ba}{\begin{eqnarray}}
\newcommand{\be}{\begin{equation}}
\newcommand{\ea}{\end{eqnarray}}
\newcommand{\ee}{\end{equation}}

\newcommand{\dt}{{\mathrm{dot}}}
\newcommand{\ld}{{\mathrm{leads}}}
\newcommand{\hy}{{\mathrm{hyb}}}
\newcommand{\tr}{\mathrm{tr}}

\newcommand{\Lv}{\mathcal{L}}
\renewcommand{\P}{\mathcal{P}}
\newcommand{\Q}{\mathcal{Q}}
\newcommand{\rrangle}{\rangle\!\rangle}
\newcommand{\llangle}{\langle\!\langle}

\newcommand{\ignore}[1]{}

\begin{document}

\title{Tunneling through molecules and quantum dots: master-equation approaches}

\author{Carsten Timm}
\email{ctimm@ku.edu}
\affiliation{Department of Physics and Astronomy, University of Kansas, Lawrence,
Kansas 66045, USA}

\date{January 7, 2008}

\begin{abstract}
An important class of approaches to the description of electronic transport
through molecules and quantum dots is based on the master equation. We discuss
various formalisms for deriving a master equation and their interrelations. It
is shown that the master equations derived by Wangsness, Bloch, and Redfield
and by K\"onig \textit{et al}.\ are equivalent. The roles of the
large-reservoir and Markov approximations are clarified. The Markov
approximation is traced back to nonzero bias voltage and temperature, whereas
interactions and the corresponding rapid relaxation in the leads are shown to be
irrelevant for the transport under certain conditions. It is explained why the
\textit{T}-matrix formalism gives incomplete results except for diagonal
density operators and to second order in the tunneling amplitudes. The
time-convolutionless master equation is adapted to tunneling problems and a
diagrammatic scheme for generating arbitrary orders in the tunneling amplitudes
is developed.
\end{abstract}



\maketitle

\section{Introduction}
\label{sec.intro}

Most approaches employed for the description of tunneling through molecules and
quantum dots fall into one of two conceptual classes: In the first, one focuses
on the dynamics of individual \emph{electrons} tunneling through the system.
Their dynamics is often described with the help of single-particle
nonequilibrium Green functions.\cite{MeW92,GRN07} This approach is sometimes
combined with \textit{ab-initio} methods\cite{SaR06,SKG06} or perturbation
theory for the interactions on the dot. In the second, one focuses on the time
evolution of the \emph{many-particle state} of the dot and integrates out the
effect of the leads. This allows one to treat the strong interactions on the
dot exactly. Typically, the reduced density operator of the dot is considered.
Its equation of motion is the master equation (ME). The two approaches become
equivalent in the absence of interactions.


The ME approach involves two steps. First, one derives the ME
from the von Neumann equation for the full system. It tells one how
the reduced density operator changes, based on its present and often its past
values. It is clear that this requires additional assumptions, since the dot is
coupled to the leads, which can be in any state. It is of course desirable to
get by with only weak assumptions.

Second, one solves the ME to find the time evolution or the
stationary state. Finding the time evolution is more complicated if the ME
contains the history of the dot. It is thus desirable to obtain a
ME that is local in time. There exist both approximate and exact
methods for achieving this.

The ME approach comes in a number of flavors, among them the
original Wangs\-ness-Bloch-Red\-field (WBR) approach,\cite{WaB53,Blo57,Red65}
the superoperator formalism, the \textit{T}-matrix formalism, the
Keldysh-contour formulation of K\"onig \textit{et
al.},\cite{ScS94,KSS96b,KSS97,THK03} and the so-called time-convolutionless
(TCL) ME.\cite{ToM75} The purpose of this paper is to clarify the
interrelations between these different pictures and to analyze some of their
problems. It is hoped that this will facilitate the comparison between results
obtained with different methods. In addition, the TCL formalism is
generalized to the tunneling case and is argued to provide a powerful tool for
studying the dynamics of a dot under bias.


\section{The Wangsness-Bloch-Redfield master equation}
\label{sec.WBR}

\subsection{Conventional derivation}
\label{sub.WBR}

This approach\cite{WaB53,Blo57,Red65} is commonly described in
text\-books,\cite{Lou74,Blu96,BrP02,GaZ04} though not in relation to particle
transport. Several groups have recently applied it to tunneling through
molecules.\cite{MAM04,ElT05,HEM06,LeL07,GoL07,MaG07,SNO07} We start from a
Hamiltonian $H = H_\dt + H_\ld + H_\hy$, where the terms describe the dot, the
leads, and hybridization between  them, respectively. $H_\dt$ may contain
vibrational or spin degrees of freedom and their coupling to the electrons. The
time evolution of the density operator $\rho$ of the \emph{full} system is
described by the von Neumann equation $\dot\rho = -i\, [H,\rho]$, where
$\hbar=1$. We wish to find the ME for the reduced density operator
$\rho_\dt(t) \equiv \tr_\ld\, \rho(t)$, where the trace is over many-particle
states of the leads.

The central assumption is that $H_\hy$ can be treated perturbatively.
Operators $A$ are transformed into the interaction picture with respect to
$H_\hy$,
\be
A_I(t) = e^{i(H_\dt+H_\ld)t} A(t)\, e^{-i(H_\dt+H_\ld)t} .
\ee
The density operator in the interaction picture satisfies
the equation
\be
\dot\rho_I = -i\, [H_{\hy,I}, \rho_I] .
\label{GZ.drho.3}
\ee
Integrating this equation from $t_0$ to $t$ yields
\be
\rho_I(t) = \rho_I(t_0) - i \int_{t_0}^t dt'\,
  [H_{\hy,I}(t'),\rho_I(t')] .
\label{GZ.rhoI.2}
\ee
Inserting this again into Eq.~(\ref{GZ.drho.3}) one obtains
\ba
\dot\rho_I(t) & = & -i\, [H_{\hy,I}(t),\rho_I(t_0)] \nonumber \\
& & {}- \int_{t_0}^t dt'\,
   \left[H_{\hy,I}(t),\left[H_{\hy,I}(t'),\rho_I(t')\right]\right] .\quad
\label{GZ.rhoI.3}
\ea
Continuing the iteration, one generates equations containing arbitrary powers
of $H_\hy$. However, in all of them all terms except for the one with the
highest power contain $\rho_I$ only at time $t_0$. This is problematic when we
introduce approximations. For example, if we describe cotunneling (fourth
order), sequential tunneling (second order) would only appear in the dependence
on the initial conditions. Section~\ref{sub.Markov} explains
how to avoid this.

We now consider the initial condition that the system was in a product state
at an early time $t_0$,
\be
\rho(t_0) = \rho_\dt(t_0) \otimes \rho_\ld^0 ,
\label{GZ.initial.2}
\ee
with $\rho_\ld^0$ describing separate thermal equilibria of the two leads;
their chemical potentials and temperatures can be different. A product state is
equivalent to dot and leads being statistically independent at time $t_0$,
which is natural if $H_\hy$ is switched on at time $t_0$. The initial
condition (\ref{GZ.initial.2}) implies that
\be
\tr_\ld H_{\hy,I}(t)\,\rho_I(t_0) = 0 ,
\label{GZ.trHr.2}
\ee
since $H_\hy$ is a sum of terms containing a single lead-electron operator.
Thus each term changes the lead electron number and thus gives zero
under the trace, since $\rho_\ld^0$ only contains states with sharp electron
number.

The trace over lead states of Eq.~(\ref{GZ.rhoI.3}) is
\be
\dot\rho_{\dt,I}(t) = - \int_{t_0}^t dt'\, \tr_\ld
   \left[H_{\hy,I}(t),\left[H_{\hy,I}(t'),\rho_I(t')\right]\right] .
\label{GZ.drho.5}
\ee
The first term in Eq.~(\ref{GZ.rhoI.3}) drops out due to
Eq.~(\ref{GZ.trHr.2}). Up to this point, the results are exact.


The integral in every term in Eq.~(\ref{GZ.drho.5}) is of the form
\be
\pm \prod A_{\dt,I} \left( \tr_\ld \prod B_{\ld,I} \rho_I \right)
  \prod\nolimits^\prime
A_{\dt,I} ,
\label{GZ.form.2}
\ee
where the first and last factors are products of zero or more dot electron
operators and $\prod B_{\ld,I}$ is a product of two lead electron operators.
The operators may have different time arguments. At this point it is usually
argued that the tunneling should have negligible effect on the leads, since they
form a \emph{large reservoir}.\cite{footnote.Born} Therefore, in any term
one makes the replacement\cite{WaB53,Blo57,Blu96,GaZ04}
\ba
\lefteqn{ \tr_\ld \prod B_{\ld,I} \rho_I \approx
  \tr_\ld \prod B_{\ld,I} \rho_{\dt,I} \otimes \rho_\ld^0 } \nonumber \\
& & = \rho_{\dt,I} \otimes \tr_\ld \prod B_{\ld,I} \rho_\ld^0 .
\hspace*{7em}
\label{GZ.appr.2}
\ea
Here, one replaces any two-time correlation function of the leads by the
correlation function in equilibrium. Gardiner and
Zoller\cite{GaZ04} point out that one only has to make this assumption
in the second-order terms.

In fact, we \emph{must} only make it in the second-order terms: If we were to
argue that since tunneling should have negligible effect on the leads we can
replace $\rho(t)$ by $\rho_\dt(t) \otimes \rho_\ld^0$ globally in  the von
Neumann equation, we get trivial results. By taking the lead trace, we obtain
$\dot\rho_\dt = -i\, \tr_\ld [H,\rho_\dt\otimes \rho_\ld^0] = -i\, \tr_\ld
[H_\dt,\rho_\dt]\otimes \rho_\ld^0$, which is just the unperturbed time
evolution. This is exact if $\rho(t) = \rho_\dt(t) \otimes \rho_\ld^0$ holds,
but this is not very useful, since the condition is generally not satisfied at
any other time. We will see that in the superoperator approach we do not have
to worry about this, since we only assume a product state at an initial time
$t_0$, as in Eq.~(\ref{GZ.initial.2}). Furthermore, in the TCL approach we can
avoid even this assumption.

If approximation (\ref{GZ.appr.2}) holds, Eq.~(\ref{GZ.drho.5}) becomes
\ba
\dot\rho_{\dt,I}(t) & = & - \int_{t_0}^t dt'\, \tr_\ld\,
   \big[H_{\hy,I}(t),\big[H_{\hy,I}(t'), \nonumber \\
& & \quad \rho_{\dt,I}(t') \otimes \rho_\ld^0 \big]\big] .
\label{GZ.drho.7}
\ea
This ME is non-local in time.



To make it local, one usually introduces the Markov approximation, which
replaces $\rho_{\dt,I}(t')$ by $\rho_{\dt,I}(t)$. This means that the rate of
change of $\rho_{\dt,I}$ at time $t$ is determined by $\rho_{\dt,I}$ at the
same time $t$ only. The approximation is usually motivated by an argument of
the following type:\cite{Blu96,GaZ04} Eq.~(\ref{GZ.drho.7}) contains two-time
correlation functions for the leads of the form (\ref{GZ.appr.2}). These
correlation functions decay rapidly on the time scale of the dot dynamics so
that they can be replaced by $\delta$-functions. We come back to this point in
Sec.~\ref{sub.Markov}.

The same assumption also implies that as long as $t-t_0$ is large compared to
the lead correlation time, one can replace $t_0$ by $-\infty$. 
With $t' = t-\tau$ one obtains
\ba
\dot\rho_{\dt,I}(t) & = & - \int_0^{\infty} d\tau\, \tr_\ld\,
   \big[H_{\hy,I}(t),\big[H_{\hy,I}(t-\tau), \nonumber \\
& & \quad \rho_{\dt,I}(t) \otimes \rho_\ld^0 \big]\big] .
\label{GZ.drho.9}
\ea
Transforming back into the Schr\"odinger picture using $\rho_{\dt,I}(t) = e^{i
H_\dt t} \rho_\dt(t) e^{-i H_\dt t}$ one finds
\ba
\lefteqn{ \dot\rho_\dt(t) = -i\, [H_\dt,\rho_\dt(t)]
- \int_0^{\infty} d\tau\, \tr_\ld\, \Big[H_\hy, } \nonumber \\
& & \!\!\Big[ e^{-i(H_\dt+H_\ld)\tau}\! H_\hy e^{i(H_\dt+H_\ld)\tau}\!\!,
   \rho_\dt(t) \otimes \rho_\ld^0 \Big]\Big] , \nonumber \\
& &
\label{GZ.drho.11}
\ea
which is local in time. The first term describes the unperturbed
time evolution of $\rho_\dt$ and the second is a correction of second
order in $H_\hy$. The restriction to second order entered
when we made the large-reservoir
approximation after iterating the equation of motion to second order.
More explicit expressions are given in App.~\ref{app.explicit}. They also show
that, if the tunneling amplitudes do not depend on wave vector, the leads only
enter through their density of states, temperature, and chemical potentials,
regardless of interactions in the leads.



At this point it is often assumed that the off-diagonal components of
$\rho_\dt$ decay rapidly and can be neglected. For some components this can be
motivated by superselection rules:\cite{WWW52,Zur82,GKZ95} If two dot states
$|m)$, $|n)$ differ in an observable that couples strongly to the environment,
unavoidable interactions lead to rapid decay of superpositions of these
states and thus of $\rho^\dt_{mn}$.\cite{Zur82} The standard
example is the charge.\cite{GKZ95,MaG07} Due to Gauss' law, the effect
of the charge can in principle be measured equally well on any arbitarily large
surface surrounding the system.\cite{GKZ95} Therefore, superpositions of dot
states with different charge are not observed.

On the other hand, the description of spin precession\cite{Blo46} requires the
off-diagonal components. Different spin states also differ in their long-range
(dipole) fields, but these fall off more rapidly than the Coulomb field. This
suggests that Gauss' law\cite{GKZ95} is crucial for superselection rules and
not just any algebraic decay.

If all off-diagonal components decay rapidly, one is left with the diagonal
components $P_m \equiv \rho^\dt_{mm}$, i.e., the probabilities of dot
many-particle states $|m)$. The principal-value terms in Eq.~(\ref{GZ.drho.13})
then cancel and one obtains
\ba
\dot P_m & = & - 2\pi \sum_{ij} \sum_p
  \big|\llangle i|(m| H_\hy |p) |j\rrangle\big|^2 \nonumber \\
& & {}\times  ( W_i\, P_m - W_j\, P_p )
  \, \delta(E_p+\epsilon_j-E_m-\epsilon_i) .\qquad
\ea
Here, $W_i\equiv \llangle i|\rho_\ld^0|i\rrangle$ is the probability to
find the leads in state $|i\rrangle$.
Defining the transition rates
\ba
R_{n\to m} & \equiv & 2\pi \sum_{ij} W_j \,
  \big|\llangle i|(m| H_\hy |n) |j\rrangle\big|^2 \nonumber \\
& & {}\times \delta(E_p+\epsilon_j-E_m-\epsilon_i) ,
\label{rate.rate.1}
\ea
we obtain the well-known \emph{rate equations}
\be
\dot P_m = \sum_n R_{n\to m} P_n - \sum_n R_{m\to n} P_m .
\label{rate.rate.3}
\ee
The first term describes transitions from other states $|n)$ to state $|m)$,
whereas the second describes transition out of state $|m)$. The rate equations
imply \emph{local} conservation of probability---$P_m$ only changes due to
probability flowing into or out of state $|m)$. This conservation law can be
implemented in a gauge theory.\cite{Tim07}

\subsection{Discussion of the Markov approximation}
\label{sub.Markov}

The Markov approximation is usually motivated by rapid decay of the lead
correlation functions.\cite{Blu96,GaZ04,StB04} In the second-order
approximation, each nonvanishing term contains one creation operator
$a^\dagger$ and one annihilation operator $a$. The result is non-zero only if
both belong to the same single-particle state. The trace $\tr_\ld$ over lead
many-particle states is replaced by a sum over single-particle states
characterized by lead index $\alpha$, wave vector $\mathbf{k}$, and spin
$\sigma$. As discussed in App.~\ref{app.explicit}, the correlation functions
are Green functions $G^{<}$, $G^{>}$. If the leads are normal metals, these
decay on the time scale of the quasiparticle lifetime. (The non-quasiparticle
background in the spectral function is broader than the quasiparticle peak,
corresponding to \emph{faster} processes, which are less critical for the
validity of the Markov approximation.) However, the quasiparticle lifetime
becomes long at the low temperatures at which experiments are performed. Does
the Markov approximation break down in the experimental temperature range? This
question is also relevant because the lead Hamiltonian $H_\ld$ used in actual
calculations does not contain any interactions. Thus in our model, there is not
broadening of the quasiparticle peak and $G^{<}$, $G^{>}$ do not decay in time.


While each term separately does not decay, their sum does. We replace the sum
over $\mathbf{k}$ by an integral over energy, including the density of states.
At low temperatures we can restrict the integral to the energy window between
the two chemical potentials $\mu_{<}$, $\mu_{>}$. Assuming a constant density
of states and $\mathbf{k}$-independent tunneling amplitudes, we end up
with integrals of the type
\be
\int_{\mu_{<}}^{\mu_{>}} dE\, e^{\pm i E \tau}
= \pm\frac{e^{\pm i \mu_{>} \tau}-e^{\pm i\mu_{<} \tau}}{i\tau} .
\ee
This expression contains a typical time scale $h/eV \equiv \tau_\ld$ for the
decay of correlations, restoring Planck's constant for the moment. Thus the
energy governing the decay of correlations is the bias, not the
electron-electron interaction. The same energy scale determines
dephasing, i.e., the decay of superpositions due to different chemical
potentials in the leads.\cite{MiM07} For arbitary temperatures, the limits of
integration are roughly $\mu_{<}-k_BT$ and $\mu_{>}\!+k_BT$ and the
characteristic time is the smaller of $h/eV$ and $h/k_BT$. Note that the
contribution from the quasiparticle lifetime is proportional to\cite{BrF04}
$1/T^2$ and is thus irrelevant at low temperatures.

For weak tunneling, or specifically if the conductance is small
compared to the quantum conductance,
\be
I/V \ll e^2/h ,
\label{cond.2}
\ee
the typical time between two tunneling events is
$\tau_0 = {e}/{I} \gg {h}/{eV} = \tau_\ld$.
Then the dot dynamics is indeed much slower than the decay of lead correlations
and the Markov approximation is justified. It  follows from the weak-tunneling
approximation, which we have to make in any case to work in low-order
perturbation theory.

This argument may fail if tunneling events are strongly
correlated.\cite{KoO05,KOA06,ElT07,LKO07} In this case, two or more tunneling
events can often take place during a time much shorter than $\tau_0=e/I$ and
the relation (\ref{cond.2}) does not guarantee the validity of the Markov
approximation.



A related point is seen if we proceed slightly differently in the
derivation. Starting from Eq.~(\ref{GZ.rhoI.3})
and inserting Eq.~(\ref{GZ.rhoI.2}) with renamed variables
\be
\rho_I(t') = \rho_I(t) - i \int_t^{t'} dt''\,
  [H_{\hy,I}(t''),\rho_I(t'')] ,
\ee
we obtain
\ba
\dot\rho_I(t) & = & -i\,[H_{\hy,I}(t),\rho_I(t_0)] \nonumber \\
& & {}- \int_{t_0}^t dt'\,
   [H_{\hy,I}(t),[H_{\hy,I}(t'),\rho_I(t)]] \nonumber \\
& & {}+ i \int_{t_0}^t dt' \int_t^{t'} dt''\,
   [H_{\hy,I}(t),[H_{\hy,I}(t'), \nonumber \\
& & \qquad [H_{\hy,I}(t''),\rho_I(t'')]]] ,
\label{Md.drho.2}
\ea
which is still exact. If we now restrict ourselves to the second order in
$H_\hy$, we can drop the last term. We have obtained an equation that is local
in time without invoking the Markov approximation.


We have pushed non-local terms into higher orders in $H_\hy$. Iterating the
procedure, we can achieve this to any order. All relevant terms contain
$\rho_I(t)$ instead of $\rho_I(t_0)$, which appears in the naive expansion in
Sec.~\ref{sub.WBR}. Thus we can for example derive a ME containing
sequential and cotunneling contributions.

Now we can make the large-reservoir approximation and replace  $\rho_I(t)$ by
$\rho_{\dt,I}(t)\otimes \rho_\ld^0$, as above. But here we perform this
replacement \emph{only} at time $t$. We then obtain, by tracing over the leads,
\ba
\dot\rho_{\dt,I}(t) & = &
   - \int_{t_0}^t dt'\, \tr_\ld
   [H_{\hy,I}(t),[H_{\hy,I}(t'), \nonumber \\
& & \quad \rho_{\dt,I}(t)\otimes \rho_\ld^0]] .
\label{Md.drho.3}
\ea
We finally replace $t_0$ by $-\infty$. This is a remnant of the Markov
approximation, but only if the ``true'' $t_0$ is finite. The choice of $t_0$
can be viewed  as a part of our model as opposed to the approximations employed
to solve it. Equation (\ref{Md.drho.3}) can also be obtained from a variational
principle, again only expanding in $H_\hy$ without explicit Markov
assumption.\cite{ZhR05}


We have obtained the same local ME in seemingly different ways.
The explanation is that we have not made independent approximations. The rapid
decay of the lead correlation functions follows from the assumption of weak
tunneling, which also allows us to use perturbation theory in $H_\hy$. On the
other hand, the large-reservoir (or Born\cite{footnote.Born}) approximation
$\rho_I(t) \cong \rho_{\dt,I}(t)\otimes \rho_\ld^0$ is logically independent.

\section{Superoperator formalism}
\label{sec.super}



The WBR ME can also be derived in the superoperator
formalism,\cite{Haa73,FiS90,BrP02,GaZ04} which facilitates
expansion to higher orders in $H_\hy$. We here define a superoperator as an
operator acting on the space of linear operators on the Hilbert space. The von
Neumann equation is written as $\dot\rho = -i\Lv \rho$, where
\be
\Lv = \Lv_\dt + \Lv_\ld + \Lv_\hy
\label{TCL1.Lanal.1}
\ee
is the Liouvillian defined by $\Lv_\dt\rho \equiv [H_\dt,\rho]$ etc. The
solution reads $\rho(t) = e^{-i\Lv(t-t_0)} \rho(t_0)$. We define projection
(super-) operators\cite{Nak58,Zwa60} $\P$, $\Q$ by $\P \rho(t) \equiv [\tr_\ld
\rho(t)] \otimes \rho_\ld^0$ and $\Q \equiv 1 - \P$. Note that $\P$ maps a
density operator onto one in product form with the leads in equilibrium, while
retaining the information on the dot state. Conversely, $\Q\rho$ contains the
information on the leads and the dot-lead correlations. It is easy to prove the
identities
\ba
\P\Lv_\dt & = & \Lv_\dt\P ,
\label{TCL1.PL.1} \\
\P\Lv_\ld & = & \Lv_\ld\P \; = \; 0 ,
\label{TCL1.PL.2} \\
\P\Lv_\hy\P & = & 0 .
\label{TCL1.PHhyP.1}
\ea
The last relation is essentially equivalent to Eq.~(\ref{GZ.trHr.2}).





The von Neumann equation can be split into two
equations,\cite{Haa73,FiS90,BrP02,GaZ04}
\ba
\frac{d}{dt}\,\P\rho & = & -i \P\Lv \P \rho - i \P\Lv\Q \rho ,
\label{TCL1.dPr.2}  \\
\frac{d}{dt}\,\Q\rho & = & -i \Q\Lv \Q\rho - i \Q\Lv\P \rho ,
\label{TCL1.dQr.2}
\ea
which can be solved by Laplace transformation,
\be
\tilde F(s) \equiv \int_0^\infty dt\, e^{-st}\, F(t) ,
\ee
where we have set $t_0=0$. We find
\ba
s\, \P\tilde\rho - \P\rho(0) & = & -i \P\Lv \P \tilde\rho
  - i \P\Lv\Q \tilde\rho ,
\label{TCL1.sPr.3}  \\
s\, \Q\tilde\rho - \Q\rho(0) & = & -i \Q\Lv \Q\tilde\rho
  - i \Q\Lv\P \tilde\rho .
\label{TCL1.sQr.3}
\ea
Inserting the solution of the second equation,
\be
\Q\tilde\rho = (s+i\Q\Lv)^{-1} \Q\rho(0) - i\, (s+i\Q\Lv)^{-1} 
  \Q\Lv\P \rho ,
\ee
into the first, we obtain
\ba
s\, \P\tilde\rho & = & \P\rho(0)
  - i \P\Lv \, (s+i\Q\Lv)^{-1} \Q\rho(0)
  - i \P\Lv \P \tilde\rho \nonumber \\
& & {}- \P\Lv \, (s+i\Q\Lv)^{-1} \Q\Lv\P \tilde\rho .
\label{TCL1.smst.2}
\ea
Using Eqs.~(\ref{TCL1.Lanal.1})--(\ref{TCL1.PHhyP.1}), we find
\ba
\lefteqn{ s\, \P\tilde\rho = \P\rho(0) } \nonumber \\
& & {}- i \P\Lv_\hy \, (s+i\Lv_\dt+i\Lv_\ld+ i\Q\Lv_\hy)^{-1} \Q\rho(0)
  \nonumber \\
& & {}- i \P\Lv_\dt \P \tilde\rho
\label{TCL1.smst.3} \\
& & {}- \P\Lv_\hy \,
  (s+i\Lv_\dt+i\Lv_\ld+i\Q\Lv_\hy)^{-1} \Q\Lv_\hy \P \tilde\rho .
\nonumber
\ea
We can insert another $\Q$ following $\Q\Lv_\hy$
since $\Q=\Q\Q$. This makes the resulting expression
$\Lv_\dt+\Lv_\ld+\Q\Lv_\hy\Q$ hermitian.

Transforming back into the time domain and shifting the
initial time back to $t_0$ we find
\ba
\lefteqn{ \frac{d}{dt}\,\P\rho =
  - i \P\Lv_\hy \, e^{-i(\Lv_\dt+\Lv_\ld+\Q\Lv_\hy\Q)(t-t_0)} \Q\rho(t_0) }
  \nonumber \\
& & {}- i \P\Lv_\dt \P \rho(t)
  - \P\Lv_\hy \, \int_{t_0}^t dt' \nonumber \\
& & \quad{}\times  e^{-i(\Lv_\dt+\Lv_\ld+\Q\Lv_\hy\Q)(t-t')}
  \Lv_\hy \P\rho(t') .\hspace*{3em}
\label{TCL1.Pr.10}
\ea
(We will show later that the projections $\Q$ in the exponentials remove all
reducible terms from the expansion in powers of $\Lv_\hy$.) Equation
(\ref{TCL1.Pr.10}) is an exact ME, which is non-local in time.

Starting from Eq.~(\ref{TCL1.Pr.10}), the weak-coupling limit as discussed in
Ref.~\onlinecite{GaZ04} now consists of (a)~neglecting all powers of $\Lv_\hy$
beyond the second and (b)~dropping the dependence on the initial condition for
$\Q\rho(t_0)$. (a) is just the weak-tunneling approximation of
Sec.~\ref{sub.WBR}. (b) neglects a term of \emph{linear} order in $\Lv_\hy$ and
has to be shown to be consistent. The rationale given in
Ref.~\onlinecite{GaZ04} is two-fold: First, the term in $\Q\rho(t_0)$ is a
small (of order $H_\hy$) correction to $\P\rho(t_0)$, and second, it is not
accumulated over time, being a correction to the initial conditions. These
arguments appear to be weak: $\Q\rho(t_0)$ and $\P\rho(t_0)$ lie in orthogonal
subspaces and it is not obvious that their
magnitudes can be meaningfully compared. Furthermore,
Eq.~(\ref{TCL1.Pr.10}) shows that $\Q\rho(t_0)$ does affect $\P\rho(t)$ for all
$t>t_0$, even to first order.

Dropping the dependence on $\Q\rho(t_0)$ is trivial if $\Q\rho(t_0)=0$. This is
not an approximation but an initial condition, see Sec.~\ref{sub.WBR}.

With approximations (a) and (b), Eq.~(\ref{TCL1.Pr.10}) becomes
\ba
\lefteqn{ \frac{d}{dt}\,\P\rho \cong - i \Lv_\dt\P\rho(t) } \nonumber \\
& & {}- \P\Lv_\hy \int_{t_0}^t dt'\,
  e^{-i(\Lv_\dt+\Lv_\ld)(t-t')}\, \Lv_\hy \P\rho(t') . \qquad
\label{TCL1.Pr.10a}
\ea
Inserting the definition of $\P$ and writing the Liouvillians as
commutators we obtain
\ba
\lefteqn{\frac{d}{dt}\,\rho_\dt \cong - i [H_\dt,\rho_\dt(t)] } \nonumber \\
& & {}- \int_{t_0}^t dt'\, \tr_\ld [H_\hy ,
  e^{-i(\Lv_\dt+\Lv_\ld)(t-t')} [H_\hy, \nonumber \\
& & \quad \rho_\dt(t')\otimes \rho_\ld^0 ]] .
\ea
Now for any Hamiltonian $H$ with associated Liouvillian $\Lv$ and
any operator (not superoperator) $A$ the identity
\be
e^{-i\Lv\tau} A = e^{-iH\tau} A \, e^{iH\tau}
\label{TCL1.expLid.2}
\ee
holds. In our case we thus get
\ba
\lefteqn{\frac{d}{dt}\,\rho_\dt \cong - i [H_\dt,\rho_\dt(t)]
- \int_{t_0}^t dt'\, \tr_\ld [H_\hy , } \nonumber \\
& & \quad [e^{-i(H_\dt+H_\ld)(t-t')} H_\hy\, e^{i(H_\dt+H_\ld)(t-t')},
  \nonumber \\
& & \quad e^{-i H_\dt (t-t')}
  \rho_\dt(t')e^{i H_\dt (t-t')} \otimes \rho_\ld^0 ]] .
\label{TCL1.drhodt.11}
\ea
Compare this to the WBR result, Eq.~(\ref{GZ.drho.11}). To get there, we have
to replace $e^{-i H_\dt (t-t')} \rho_\dt(t')e^{i H_\dt (t-t')}$ by
$\rho_\dt(t)$. This is \emph{nearly} the same: $\rho_\dt(t)$ is described by the
full Hamiltonian $H$, while Eq.~(\ref{TCL1.drhodt.11}) only contains
the unperturbed time evolution due to $H_\dt$. This is
consistent with the second-order approximation, since any
correction to the unperturbed time evolution adds more
powers of $H_\hy$. Thus the Markov property again follows.
We can extend the range of integration to $t_0\to
-\infty$ arguing as in Sec.~\ref{sub.Markov}.

Note that we did not need the large-reservoir assumption $\rho(t) \cong
\rho_\dt(t)\otimes \rho_\ld^0$ in this approach. We have only assumed the
density operator to be of this product form at an early time $t_0$---a much
weaker assumption.

\section{The \textit{T}-matrix approach and Fermi's Golden Rule}

The \textit{T}-matrix approach\cite{BrF04,Ake99,GoL04,KOA06,JoS06,ElT07,LKO07}
and its leading-or\-der approximation, Fermi's Golden
Rule,\cite{KOO04,RWS06,MiB07} are used by several groups to describe tunneling
processes, since they provide a straightforward derivation of the transition
rates in the \emph{diagonal} rate equations. The approach is presented in many
textbooks. Bruus and Flensberg\cite{BrF04} discuss it in relation to the
tunneling problem.


The derivation starts out by writing
\be
H(t) = \underbrace{H_\dt + H_\ld}_{=H_0}
  + \underbrace{H_\hy\,e^{\eta t}}_{=V(t)} ,
\ee
where $\eta$ is small and positive. Thus the hybridization is switched on
very slowly. We assume that the system was in an eigenstate $|i\rangle$ of $H_0$
at time $t_0$. The probability that it is in another eigenstate
$|f\rangle$ at time $t$ reads
\be
|\langle f|i(t)\rangle|^2
\equiv \left|\langle f|\,  \mathcal{T}
  \exp\left( -i \int_{t_0}^t dt'\, V_I(t') \right) \,
  |i\rangle \right|^2 ,
\ee
where $\mathcal{T}$ is the time-ordering operator.
The transition rate between states $|i\rangle$ and
$|f\rangle$ is then defined as
\be
\Gamma_{fi} \equiv \frac{d}{dt}\, |\langle f|i(t)\rangle|^2 .
\label{Tm.Gafi.5}
\ee
Taking the limit $\eta\to 0^+$ and defining the \textit{T}-matrix
\ba
\lefteqn{ T(E_i) \equiv H_\hy + H_\hy\,\frac{1}{E_i - H_0 + i0^+}\,H_\hy }
  \nonumber \\
& & {}+ H_\hy\,\frac{1}{E_i - H_0 + i0^+}\,H_\hy\,
  \frac{1}{E_i - H_0 + i0^+}\,H_\hy \nonumber \\
& & {}+ \ldots
\label{Tm.Tdef.1}
\ea
one obtains
\be
\Gamma_{fi} = 2\pi\, \delta(E_i-E_f)\, |\langle f|T|i\rangle|^2 ,
\label{Tm.FGR.2}
\ee
where $E_i$ and $E_f$ are the eigenenergies of states $|i\rangle$ and
$|f\rangle$, respectively.
The leading order is obtained by replacing $T$ by $H_\hy$,
\be
\Gamma_{fi} = 2\pi \, \delta(E_i-E_f)\, |\langle f|H_\hy|i\rangle|^2 ,
\label{Tm.FGR.4}
\ee
which is Fermi's Golden Rule.


To draw the connection with the rate equations, we choose the states
$|i\rangle$, $|f\rangle$ as product states of many-particle states of the dot,
$|m)$, and of the leads, $|i\rrangle$. Summing $\Gamma_{fi}$
over the lead states we obtain the transition rates
\be
\tilde R_{n\to m} = 2\pi \sum_{if} W_i
  \big| \llangle f| (m| T |n) |i\rrangle \big|^2
  \delta(E_n + \epsilon_i - E_m - \epsilon_f)
\label{Tm.R.4}
\ee
from state $|n)$ to $|m)$. Here, $E_m$ ($\epsilon_i$) are eigenenergies of dot
(lead) states and $W_i$ is the probability to find the leads in initial state
$|i\rangle$ at time $t_0\to-\infty$. To write down Eq.~(\ref{Tm.R.4}), one has
to make the assumption that the probability $W_i$ is independent of the state
of the dot. This means that the system is in a product state
$\rho=\rho_\dt\otimes\rho_\ld$ at time $t_0$. This is the same assumption
usually made in density-operator approaches. If the leads are in equilibrium at
time $t_0$ one can express $W_i$ in terms of Fermi functions.

In the next step, $\tilde R_{n\to m}$ is identified with the transition rate
$R_{n\to m}$ appearing in the rate equations. However, what is actually
calculated is the rate of change of the probability of state $|m)$ under the
condition that the dot was in state $|n)$ at time $t_0\to -\infty$, cf.\
Eqs.~(\ref{Tm.Gafi.5}) and (\ref{Tm.R.4}). On the other hand, in the
density-operator approach one calculates the rate of change of the probability
of state $|m)$ under the condition that it is in state $|n)$ at the \emph{same}
time $t$, immediately before a possible transition. The two are the same only
if the dot remains in state $|n)$ from time $t_0$ through $t$. This is of
course not usually the case.

In the sequential-tunneling approximation, one evaluates the rates to second
order in $H_\hy$. Since two powers of $H_\hy$ are required for the final
transition to state $|m)$, no transitions can occur between $t_0$ and $t$.
Therefore, to second order, where the \textit{T}-matrix approach reduces to
Fermi's Golden Rule, it gives the same transition rates $R_{n\to m}$ as the WBR
approach. Beyond leading order, $\tilde R_{n\to m}$ and $R_{n\to m}$ describe
different quantities.


To leading order, $T\cong H_\hy$, the rates $\tilde R_{n\to m}$ in
Eq.~(\ref{Tm.R.4}) are indeed identical to the the WBR result,
Eq.~(\ref{rate.rate.1}). The latter has been obtained under the Markov
assumption. One might wonder where the Markov assumption entered in the
\textit{T}-matrix formalism. It is implied in the derivation, since to second
order $\rho$ does not change between times $t_0$ and $t$ anyway.


The rate equations (\ref{rate.rate.3}) appear to be obvious. However, can they
be \emph{derived} in the \textit{T}-matrix framework? Certainly not without
further assumptions, since they omit the off-diagonal components of $\rho_\dt$
necessary for a complete description.


\section{The Keldysh-contour approach of K\"onig \textit{et al.}}

K\"onig \textit{et al.}\cite{ScS94,KSS96b,KSS97,THK03}\ have developed a
diagrammatic technique to generate a perturbative expansion in the tunneling
amplitudes. This approach has also been applied to tunneling through
molecules.\cite{THK03,WeB07} In the present section we show how it is
related to WBR theory. Reference \onlinecite{KSS96b} concerns a quantum dot
with electron-electron and electron-vibration interactions. A unitary
transformation replaces the latter with an exponential operator in the
tunneling Hamiltonian.\cite{LaF63} We do not consider this transformation here.
This does not restrict the models covered by our discussion---we include
any bosonic modes and electron-boson interactions into $H_\dt$.

As usual, the system is assumed to be in a product state at an
early time $t_0$ with the leads in separate equilibria. The propagator $\Pi$
of $\rho_\dt$ from time $t_0$ to $t\ge t_0$ is defined by
\be
\rho_\dt(t) = \Pi(t,t_0)\, \rho_\dt(t_0) .
\label{KSSS.4}
\ee
(We use the same notation and time ordering as elsewhere in this
paper.) $\Pi(t,t_0)$ is represented by a diagram on the Keldysh contour between
times $t_0$ and $t$, Fig.~3 in Ref.~\onlinecite{KSS96b}. K\"onig \textit{et
al}.\ then identify the irreducible part $\Sigma_K$ of $\Pi$ (a subscript is
added to distinguish $\Sigma_K$ from $\Sigma$ introduced above), defined as the
sum of all diagrams that cannot be cut at an intermediate time without cutting
a lead line representing the pairing of $a^\dagger_{\alpha\mathbf{k}\sigma}$
and $a_{\alpha\mathbf{k}\sigma}$. $\Pi$ is expressed in
terms of $\Sigma_K$ by a Dyson-type equation,
\ba
\Pi(t,t_0) & = & \Pi^{(0)}(t,t_0) + \int_{t_0}^t dt_2 \int_{t_0}^{t_2} dt_1\,
  \Pi^{(0)}(t,t_2) \nonumber \\
& & \qquad{}\times \Sigma_K(t_2,t_1)\, \Pi(t_1,t_0) ,
\label{KSSS.Dy.4}
\ea
containing the bare propagator
\be
[\Pi^{(0)}(t,t_0)]^{nn'}_{mm'} = \delta_{mm'} \delta_{nn'}
  e^{-i(E_n-E_m)(t-t_0)} .
\label{KSSS.Dy.4a}
\ee
In Ref.~\onlinecite{KSS96b}, Eqs.~(\ref{KSSS.4})--(\ref{KSSS.Dy.4a}) are
presented for arbitary initial time $t'\ge t_0$. Equation (\ref{KSSS.4}) would
then imply $\Pi(t,t'') = \Pi(t,t')\,\Pi(t',t'')$, which contradicts
Eq.~(\ref{KSSS.Dy.4}).\cite{footnote.KSSS} This is not a problem since
the equations with initial
time $t_0$ are sufficient for the derivation of the ME.

We want to find the propagator $\Pi$ explicitly. Clearly, we have
$\P\rho(t) = \P e^{-i\Lv(t-t_0)}\, \rho(t_0)$.
Since the initial condition $\Q\rho(t_0)=0$ is assumed, we find
\be
\P\rho(t) = [\P e^{-i\Lv(t-t_0)}\, \P]\, \P\rho(t_0)
  \equiv \Pi(t,t_0)\, \P\rho(t_0) .
\label{KSSS.Pdef2}
\ee
This defines the propagator $\Pi$ for initial time
$t_0$. The last factor $\P$ in $\Pi$ is expendable.

Can we find $\Sigma_K$ to satisfy Eq.~(\ref{KSSS.Dy.4})? We expand the
exponential in Eq.~(\ref{KSSS.Pdef2}), noting that  all lead creation and
annihilation operators must be paired for the result to be nonzero since
$\P\Lv_\hy\P=0$. Diagrammatically, this is represented by a lead-fermion line
connecting two $H_\hy$ insertions.\cite{KSS96b} The lead trace then gives a
Fermi factor for any pair of insertions. With regard to this pairing, we can
identify the irreducible part $\Sigma_K$ and write
\ba
\lefteqn{ \P e^{-i\Lv(t-t_0)} \P =
  \P e^{-i(\Lv_\dt+\Lv_\ld)(t-t_0)}
  + \P \int_{t_0}^t \! dt_2 \int_{t_0}^{t_2} \!\! dt_1 } \nonumber \\
& & \quad{}\times  e^{-i(\Lv_\dt+\Lv_\ld)(t-t_2)}
  \, \Sigma_K(t_2,t_1) \nonumber \\
& & \quad{}\times e^{-i(\Lv_\dt+\Lv_\ld)(t_1-t_0)} \P + \ldots \nonumber \\
& & = \P e^{-i(\Lv_\dt+\Lv_\ld)(t-t_0)}
  + \P \int_{t_0}^t dt_2 \int_{t_0}^{t_2} dt_1 \nonumber \\
& & \quad{}\times e^{-i(\Lv_\dt+\Lv_\ld)(t-t_2)}
  \, \Sigma_K(t_2,t_1)\, e^{-i\Lv(t_1-t_0)} \P .\qquad
\label{KSSS.Dy.8}
\ea
If we insert $\P$ at some intermediate time $t_n$ in
Eq.~(\ref{KSSS.Dy.8}) this forces all lead-fermion operators to be paired for
times smaller than and larger than $t_n$
separately. This is equivalent to
the diagram being reducible at time $t_n$.
Conversely, if a diagram is reducible at
time $t_n$, inserting $\P$ there does not change the result.
Consequently,
\ba
\lefteqn{ \P e^{-i\Lv(t-t_0)} \P = \P e^{-i(\Lv_\dt+\Lv_\ld)(t-t_0)}
  + \P \int_{t_0}^t \!\! dt_2 \int_{t_0}^{t_2} \! dt_1 } \nonumber \\
& & {}\times e^{-i(\Lv_\dt+\Lv_\ld)(t-t_2)}
  \, \Sigma_K(t_2,t_1)\, \P e^{-i\Lv(t_1-t_0)} \P .\qquad
\label{KSSS.Dy.9}
\ea
Here, we can identify the bare propagator
\be
\Pi^{(0)}(t,t') \equiv \P e^{-i(\Lv_\dt+\Lv_\ld)(t-t')} .
\label{KSSS.Dy.9a}
\ee
These two equations correspond to Eqs.~(\ref{KSSS.Dy.4}), (\ref{KSSS.Dy.4a})
except that we had to introduce another projection $\P$ at the final time.
Apart from this, we recover the Dyson-type equation of
Ref.~\onlinecite{KSS96b} for initial time $t_0$.

Inserting the Dyson-type equation (\ref{KSSS.Dy.9}) for the propagator into
Eq.~(\ref{KSSS.Pdef2}) and taking the time derivative, one obtains a
ME,\cite{KSS96b}
\be
\frac{d}{dt}\P\rho(t) = -i\, \Lv_\dt \, \P\rho(t)
  + \P\int_{t_0}^t dt'\, \Sigma_K(t,t')\, \P\rho(t') .
\label{KSSS.master.3}
\ee
This corresponds to Eq.~(25) in Ref.~\onlinecite{KSS96b} (a factor of $\P$ on
the left can be included into the definition of $\Sigma_K$).


\begin{figure}[tbh]
\includegraphics[scale=0.5]{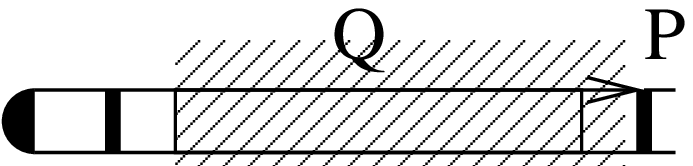}
\caption{Diagram for the second term in the ME
(\protect\ref{KSSS.master.4}). Time is increasing to the right. The bare
propagator $e^{-i(\Lv_\dt+\Lv_\ld)(t-t')}$ 
for the density operator (not occuring here) would be
represented by just two lines. The full propagator $e^{-i\Lv(t-t')}$ is shown
as a hatched box. Its irreducible part
$e^{-i\Q\Lv\Q(t-t')}$ is distinguished by writing ``Q'' beside it.
All additional projection operators are likewise indicated by ``P'' or ``Q''.
An insertion of $-i\Lv_\hy$ is denoted by a heavy bar connecting the two lines.
The projected density operator $\P\rho(t')$ at time $t'$ is shown as a filled
semicircle.}
\label{fig.rate_NTCL_Sigma}
\end{figure}

We can gain further insight by returning to Eq.~(\ref{TCL1.Pr.10}), restricted
to the case $\Q\rho(t_0)=0$,
\ba
\lefteqn{ \frac{d}{dt}\,\P\rho =
  - i\, \Lv_\dt \P \rho(t) - \P\Lv_\hy \, \int_{t_0}^t dt' } \nonumber \\
& & {}\times e^{-i(\Lv_\dt+\Lv_\ld+\Q\Lv_\hy\Q)(t-t')} \Lv_\hy\, \P\rho(t')
  .\quad
\label{KSSS.master.4}
\ea
Comparing to Eq.~(\ref{KSSS.master.3}) we find
\be
\P\Sigma_K(t,t')\P = \P\Lv_\hy
  e^{-i(\Lv_\dt+\Lv_\ld+\Q\Lv_\hy\Q)(t-t')}\! \Lv_\hy\P .
\label{KSSS.SK.4}
\ee
Thus the ME of Ref.~\onlinecite{KSS96b} is equivalent to the WBR ME to all
orders in $H_\hy$ and we have derived an explicit expression for the
irreducible part. For later, we introduce in Fig.~\ref{fig.rate_NTCL_Sigma} a
diagrammatic representation for the second term in Eq.~(\ref{KSSS.master.4}).

Now the factors of $\Q$ in the exponential find a
natural interpretation: They remove all reducible terms from the expansion of
Eq.~(\ref{KSSS.SK.4}). This is because at a point where a diagram is reducible,
one can insert a $\P$. But if a $\Q$ is present at this point, we obtain
$\P\Q=\Q\P=0$. Thus all diagrams in the expansion that are reducible
to the left or to the right of an insertion of $\Lv_\hy$ vanish.


\section{The time-convolutionless master equation}
\label{sec.TCL}


This approach leads to a ME that is local in time and exact, and
thus avoids the Markov assumption. The Markov assumption is valid for weak
tunneling and not strongly correlated tunneling events, as discussed above, but
becomes increasingly dubious at higher orders in $H_\hy$ or in resummation or
non-perturbative schemes. The approach was developed by Tokuyama and
Mori\cite{ToM75} and others\cite{HSS77,Sae88,Ahn94} and is discussed in
Ref.~\onlinecite{BrP02}. This section adapts it to the tunneling problem.

We again start from Eqs.~(\ref{TCL1.dPr.2}) and (\ref{TCL1.dQr.2}) and
express $\Q\rho(t)$ in the first equation in terms of $\P\rho(t)$
and $\Q\rho(t_0)$ with the help of the second. We do not make any assumptions
on $\Q\rho(t_0)$. The solution of Eq.~(\ref{TCL1.dQr.2}) reads
\ba
\Q\rho(t) & = & e^{-i\Q\Lv\Q(t-t_0)} \Q \rho(t_0) \nonumber \\
& & {}- i \int_{t_0}^t dt'\,
  e^{-i\Q\Lv\Q (t-t')}\, \Q\Lv\P \rho(t') .
\ea
Next, we express $\rho(t')$ by propagating the full density operator backward
in time,
\ba
\Q\rho(t) & = & e^{-i\Q\Lv\Q(t-t_0)} \Q \rho(t_0)
  - i \int_{t_0}^t dt' \, e^{-i\Q\Lv\Q (t-t')} \nonumber \\
& & \qquad{}\times \Q\Lv\P e^{-i\Lv(t'-t)}
  \big[\P\rho(t) + \Q\rho(t)\big] .
\ea
Moving all terms in $\Q\rho(t)$ to the left we obtain
\be
( 1 - \Sigma )\, \Q\rho(t) = e^{-i\Q\Lv\Q(t-t_0)} \Q \rho(t_0)
  + \Sigma\, \P\rho(t)
\label{TCL1.Qr.3}
\ee
with the superoperator
\be
\Sigma(t-t_0) \equiv -i \int_{t_0}^t dt'\,
  e^{-i\Q\Lv\Q (t-t')} \Q\Lv\P e^{-i\Lv(t'-t)} .
\label{TCL1.Si.2}
\ee
The time argument $t-t_0$ will be suppressed if confusion is unlikely.
The integral can be performed, giving
\ba
\Sigma(t-t_0) & = & \int_{t_0}^t dt'\, \bigg(
  e^{-i\Q\Lv\Q(t-t')} \Q\, \frac{\partial}{\partial t'}\,
  e^{-i\Lv (t'-t)} \nonumber \\
& & {}+ \left[ \frac{\partial}{\partial t'}\,
  e^{-i\Q\Lv\Q (t-t')} \right] \Q e^{-i\Lv(t'-t)} \bigg) \nonumber \\
& = & \Q - e^{-i\Q\Lv\Q (t-t_0)} \Q e^{-i\Lv (t_0-t)} .
\label{TCL1.Si.5}
\ea
The integral form (\ref{TCL1.Si.2})
is more suitable for the expansion in $\Lv_\hy$, though.

Applying the inverse $(1-\Sigma)^{-1}$ to Eq.~(\ref{TCL1.Qr.3}), we obtain
\be
\Q\rho(t) = (1-\Sigma)^{-1} e^{-i\Q\Lv\Q(t-t_0)} \Q \rho(t_0)
  + (1-\Sigma)^{-1} \Sigma\, \P\rho(t) .
\label{TCL1.Qr.5}
\ee
This remarkable equation asserts that we can reconstruct $\Q\rho$ and
thus $\rho=\P\rho+\Q\rho$ at time $t$ from $\P\rho$ at time
$t$ and $\Q\rho$ at some arbitrarily early time $t_0$, even though
$\P\rho$ only contains information on the dot state.


Inserting Eq.~(\ref{TCL1.Qr.5}) into Eq.~(\ref{TCL1.dPr.2}) we obtain an
equation of motion for $\P\rho$ alone,
\ba
\frac{d}{dt}\,\P\rho(t) & = & - i \P\Lv(1-\Sigma)^{-1} \P\rho(t) \nonumber \\
& & {}- i\P\Lv(1-\Sigma)^{-1} e^{-i\Q\Lv\Q(t-t_0)} \Q\rho(t_0) ,\qquad
\label{TCL1.Pr.6}
\ea
which, together with Eq.~(\ref{TCL1.Si.2}) or (\ref{TCL1.Si.5}),
constitutes the TCL ME. It indeed only contains $\P\rho$ at the
time $t$. Eqs.~(\ref{TCL1.Pr.6}) and (\ref{TCL1.Si.2})
can be generalized for time-dependent Hamiltonians\cite{Sae88,Ahn94} by
replacing the multiplication with $t-t_0$ by a time integral.

The TCL ME is exact, but relies on the condition that the inverse
of $1-\Sigma$ exists. For a different system not involving
transport, one can find cases when $1-\Sigma$ is singular.\cite{BKP99} Since
$\Sigma$ vanishes at $t=t_0$, cf.\ Eq.~(\ref{TCL1.Si.5}), Breuer \textit{et
al.}\cite{BKP99,BrP02} conclude that $1-\Sigma$ is not singular for
sufficiently small $t-t_0$. This does not directly apply to our case, since
the time dependence is governed by the dynamics of the full system, which
permits arbitarily large excitation energies. In other words, the eigenvalues
of $\Lv$ and $\Q\Lv\Q$ in Eq.~(\ref{TCL1.Si.5}) are not bounded.

Using Eqs.~(\ref{TCL1.Lanal.1})--(\ref{TCL1.PHhyP.1}) we
obtain a more explicit form,
\ba
\lefteqn{ \frac{d}{dt}\,\P\rho(t) = - i \Lv_\dt \P\rho(t)
  - i \P\Lv_\hy \, (1-\Sigma)^{-1} \P\rho(t) } \nonumber \\
& & {}- i \P\Lv_\hy \, (1-\Sigma)^{-1}\,
  e^{-i(\Lv_\dt+\Lv_\ld+\Q\Lv_\hy\Q)(t-t_0)} \nonumber \\
& & \quad{}\times \Q\rho(t_0) \hspace*{5em}
\label{TCL1.Pr.8}
\ea
with
\ba
\Sigma(t-t_0) & = & - i \Q \int_{t_0}^t dt'\,
  e^{-i (\Lv_\dt+\Lv_\ld+\Q\Lv_\hy\Q)(t-t')} \nonumber \\
& & {}\times \Lv_\hy\P e^{-i\Lv(t'-t)} .
\label{TCL1.Si.8}
\ea
We have used that $\P(1-\Sigma)^{-1} = \P$, as can be seen by expanding
in powers of $\Sigma$. Equation (\ref{TCL1.Pr.8}) should be compared to the
non-local ME (\ref{TCL1.Pr.10}).


For the case $\Q\rho(t_0)=0$, the equation simplifies to
\be
\frac{d}{dt}\,\P\rho(t) = - i \Lv_\dt \P\rho(t)
  - i \P\Lv_\hy [1-\Sigma(t-t_0)]^{-1} \P\rho(t) .
\label{TCL1.Pr.9}
\ee
It is then tempting to take the limit $t_0\to -\infty$. One
has to check whether this limit exists for $\Sigma(t-t_0)$.

As advertized, the TCL ME (\ref{TCL1.Pr.8}) is local in time,
although the dynamics is generally not Markovian. The memory effects have been
shifted into the time dependence of the coefficients of
$\P\rho$.\cite{RoP97,MIP04,Whi07} This works because the integro-differential
WBR ME is linear. One can then show that a purely differential
equation with the same solution exists.\cite{RoP97} Maniscalco \textit{et
al.}\cite{MIP04} use the exact solution for the damped harmonic oscillator to
illustrate that non-Markovian dynamics is indeed compatible with a TCL
formulation. This and related results\cite{Whi07,PeB06} do not involve
transport.


We briefly comment on the question of positivity of the reduced density
operator, i.e., the requirement that all its eigenvalues are non-negative. As
in Ref.~\onlinecite{Whi07}, the coefficients in the TCL ME are
time-dependent. Thus Lindblad's\cite{Lin76} criterion for positivity does not
apply. However, the TCL ME is exact so that its solution for
$\P\rho$ satisfies $\P\rho=[\tr_\ld \rho] \otimes \rho_\ld^0$ at all times and
thus certainly satisfies positivity. It is an important question whether
perturbative approximations destroy this property.


To obtain the sequential-tunneling approximation to Eq.~(\ref{TCL1.Pr.8}), we
expand to second order in $\Lv_\hy$,
\begin{widetext}
\ba
\lefteqn{ \frac{d}{dt}\,\P\rho(t) \cong
  - i \Lv_\dt \P\rho(t)
  - \P\Lv_\hy  \int_{t_0}^t dt'\,
  e^{-i (\Lv_\dt+\Lv_\ld)(t-t')} \Lv_\hy e^{-i(\Lv_\dt+\Lv_\ld)(t'-t)}
  \P\rho(t) } \nonumber \\
& & {}- i \P\Lv_\hy
  e^{-i(\Lv_\dt+\Lv_\ld)(t-t_0)} \Q\rho(t_0)
- \P\Lv_\hy \int_{t_0}^t dt'\,
  e^{-i(\Lv_\dt+\Lv_\ld)(t-t')} \Lv_\hy
  e^{-i(\Lv_\dt+\Lv_\ld)(t'-t_0)}\, \Q\rho(t_0) .\quad
\label{TCL1.master2.5}
\ea
\end{widetext}
The first term describes the unperturbed time evolution. The two
inhomogeneous terms describe the effect of a deviation of the state
at time $t_0$ from a product state with leads in equilibrium. The
third term is the only one of first order in the tunneling amplitudes, thus for
$\Q\rho(t_0)=0$ there are no first-order terms. For readers familiar with
optical response theory this may seem surprising. We briefly discuss
first-order terms in App.~\ref{app.first}.


\subsection{Perturbative expansion}

Since expansions beyond second order are clearly cumbersome to write down, a
diagrammatic representation is helpful. We here assume the simplifying
initial condition $\Q\rho(t_0)=0$. We first expand Eq.~(\ref{TCL1.Pr.9}) in
powers of $\Sigma$,
\ba
\lefteqn{ \frac{d}{dt}\,\P\rho(t) = - i \Lv_\dt \P\rho(t) } \nonumber \\
& & {}- i \P\Lv_\hy \, (\Sigma + \Sigma\Sigma + \Sigma\Sigma\Sigma + \ldots)
  \P\rho(t) .
\label{TCL1.master.10}
\ea
The first-order term vanishes since $\P\Lv_\hy\P=0$. Note that the series
$1+\Sigma+\Sigma^2+\ldots$ might not converge even if the inverse of $1-\Sigma$
exists.

\begin{figure}[tbh]
\includegraphics[scale=0.5]{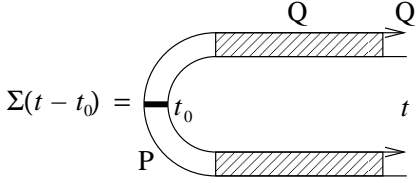}
\caption{Diagram for the superoperator $\Sigma(t-t_0)$. The interpretation
of symbols is given in the caption of Fig.~\protect\ref{fig.rate_NTCL_Sigma}.
The right-most superoperator in Eq.~(\protect\ref{TCL1.Si.8})
corresponds to the lower right corner of the diagram.}
\label{fig.Sigma_full}
\end{figure}

\begin{figure}[tbh]
(a)\quad\includegraphics[scale=0.5]{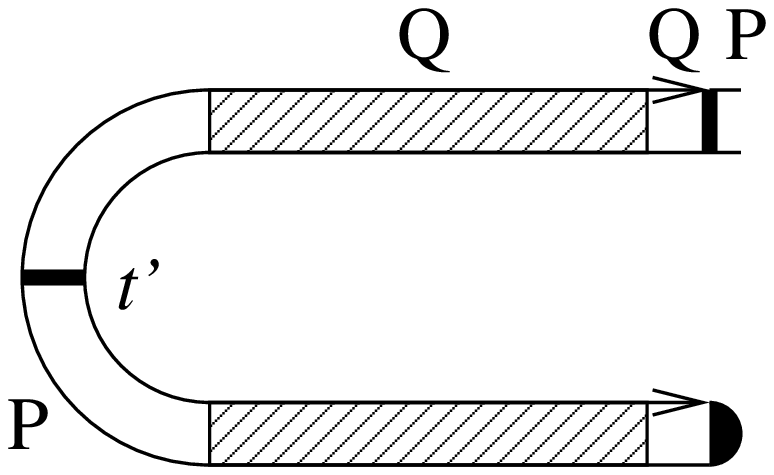}

(b)\quad\includegraphics[scale=0.5]{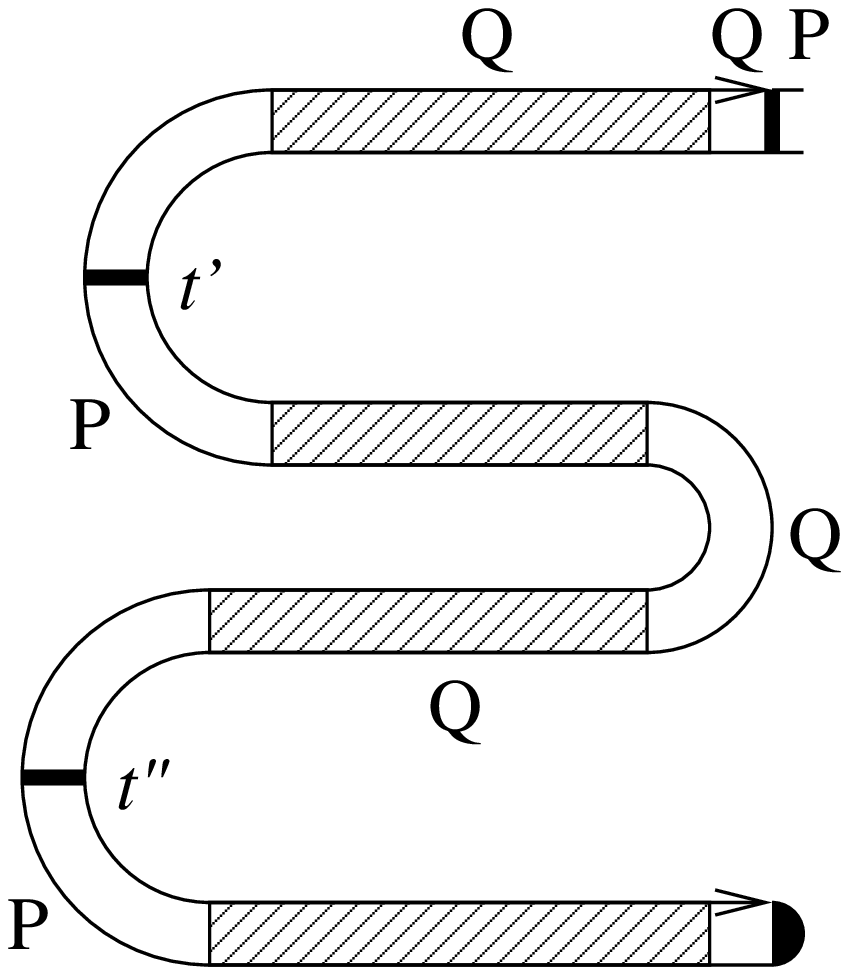}
\caption{Contributions to the TCL ME (\protect\ref{TCL1.master.10})
containing (a) one and (b) two powers of the superoperator $\Sigma$. The
filled semicircle denotes the density operator $\P\rho(t)$.
In (b), no time ordering of $t'$ and $t''$ is implied.}
\label{fig.rate_nSigma}
\end{figure}

The superoperator $\Sigma$, Eq.~(\ref{TCL1.Si.8}), first propagates the density
operator backward in time, projects it, inserts a perturbation $\Lv_\hy$, and
then propagates it forward again. Its diagrammatic representation is shown in
Fig.~\ref{fig.Sigma_full}. The contributions to Eq.~(\ref{TCL1.master.10}) with
one and two powers of $\Sigma$ are represented by the diagrams in
Fig.~\ref{fig.rate_nSigma}. It is obvious how the series continues.

\begin{figure}[tbh]
\includegraphics[scale=0.5]{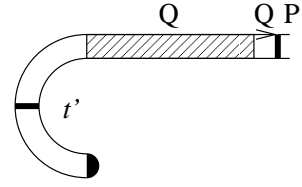}
\caption{General form of all terms involving tunneling in the TCL ME
(\protect\ref{TCL1.master.10}). Here, the filled semicircle
denotes the projected density operator $\P\rho$ at time $t'$.}
\label{fig.rate_TCL_trunc}
\end{figure}

This is a good place to compare to the approach of WBR and K\"onig \textit{et
al.} The contribution from the tunneling is in this case given by the last term
in Eq.~(\ref{KSSS.master.4}) or the diagram Fig.~\ref{fig.rate_NTCL_Sigma}. To
show its consistency with the TCL equation, we note that all tunneling
contributions in the TCL approach are of the form shown in
Fig.~\ref{fig.rate_TCL_trunc}. This is indeed just a deformation of
Fig.~\ref{fig.rate_NTCL_Sigma} (the ``Q'' adjacent to the final $\Lv_\hy$ is
expendable). The upper part of $\Sigma$ is thus identical to the irreducible
part $\Sigma_K$ without the final $\Lv_\hy$. This is also seen by comparing the
algebraic expressions (\ref{KSSS.SK.4}) and (\ref{TCL1.Si.8}).



In order to expand the ME in powers of
$\Lv_\hy$, we next expand the exponentials as
\ba
\lefteqn{ e^{-i(\Lv_\dt+\Lv_\ld+\Q\Lv_\hy\Q)(t-t')}
  = e^{-i(\Lv_\dt+\Lv_\ld)(t-t')} } \nonumber \\
& & {}- i\,\Q \int_{t'}^t dt_1\, e^{-i (\Lv_\dt+\Lv_\ld)(t-t_1)} \Lv_\hy\Q
  \nonumber \\
& & \quad{}\times e^{-i (\Lv_\dt+\Lv_\ld)(t_1-t')} \nonumber \\
& & {}- \Q \int_{t'}^t dt_2 \int_{t'}^{t_2} dt_1\,
  e^{-i (\Lv_\dt+\Lv_\ld)(t-t_2)}
  \Lv_\hy\Q \nonumber \\
& & \quad{}\times e^{-i(\Lv_\dt+\Lv_\ld)(t_2-t_1)} \Lv_\hy\Q \nonumber \\
& & \quad{}\times e^{-i(\Lv_\dt+\Lv_\ld)(t_1-t')} + \ldots\,
\ea
and
\ba
\lefteqn{ e^{-i(\Lv_\dt+\Lv_\ld+\Lv_\hy)(t'-t)}
  = e^{-i(\Lv_\dt+\Lv_\ld)(t'-t)} } \nonumber \\
& & {}+ i \int_{t'}^t dt_1\, e^{-i (\Lv_\dt+\Lv_\ld)(t'-t_1)} \Lv_\hy
  \nonumber \\
& & \quad{}\times e^{-i (\Lv_\dt+\Lv_\ld)(t_1-t)}  + \ldots\,
\label{TCL1.Uexp.4}
\ea
The odd terms in Eq.~(\ref{TCL1.Uexp.4}) obtain an additional
minus sign, since the reversed time order
gives an additional minus sign for each integral.

\begin{figure}[tbh]
(a)\includegraphics[scale=0.5]{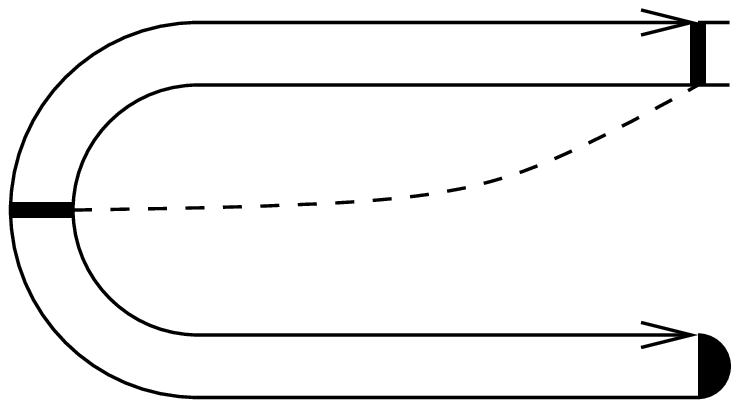}

(b)\includegraphics[scale=0.5]{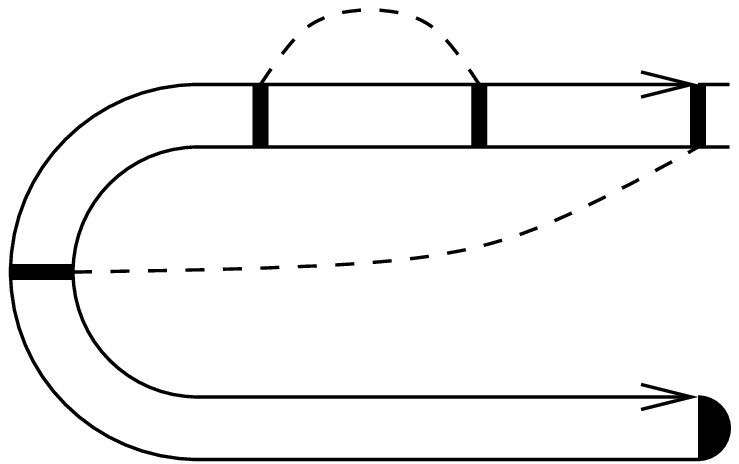}

(c)\includegraphics[scale=0.5]{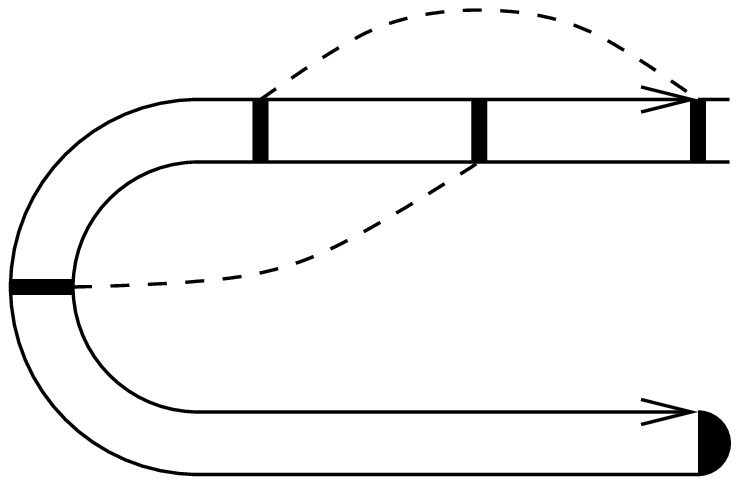}

(d)\includegraphics[scale=0.5]{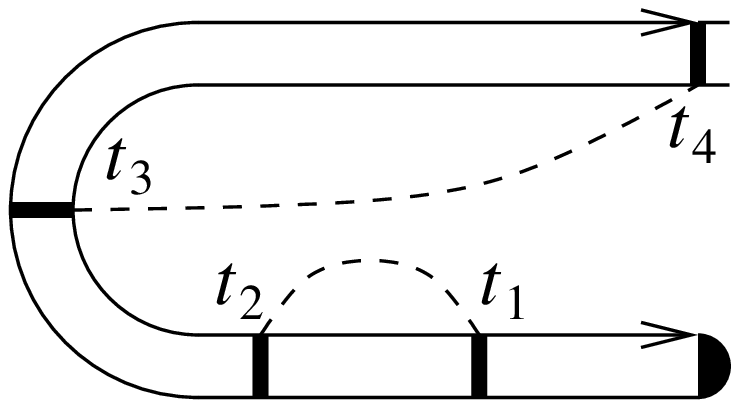}

(e)\includegraphics[scale=0.5]{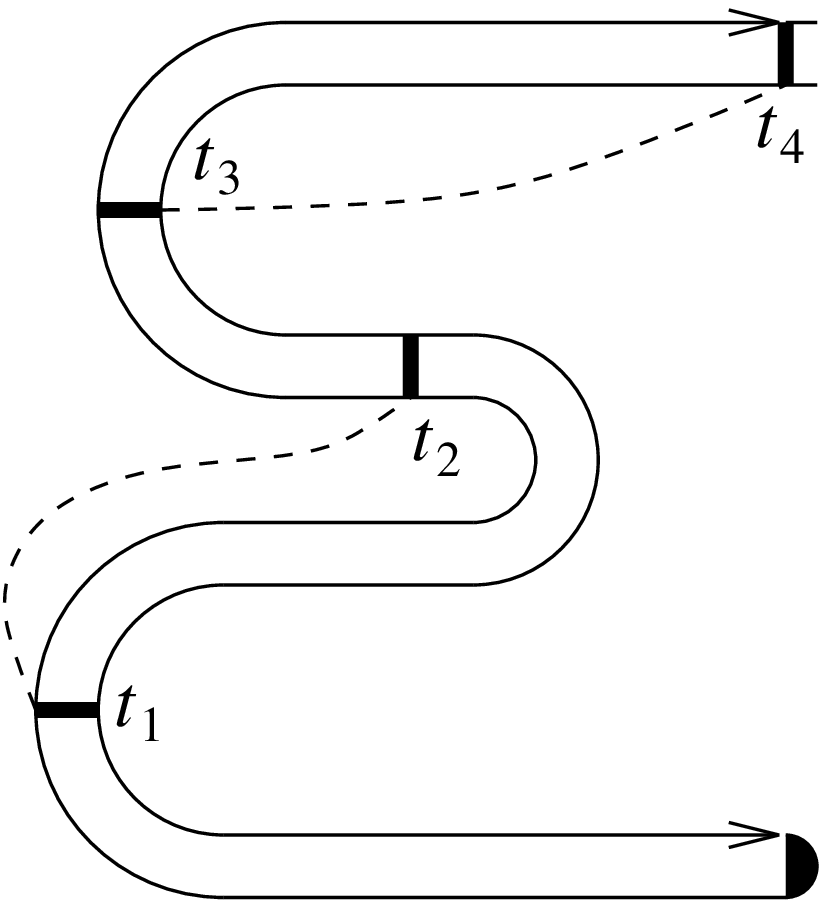}
\caption{Diagrams for all terms in the TCL ME up to fourth order in the
tunneling amplitudes. The dashed lines denote the pairing of lead-electron
operators $a$, $a^\dagger$ with the same quantum numbers.}
\label{fig.rate_nSm}
\end{figure}


As discussed earlier, any insertion of $\P$ forces an expression to be
reducible at that point and any insertion of $\Q$ means that the diagram must
not be reducible at that point. $\P$ and $\Q$ thus govern the construction of
non-vanishing diagrams but do not affect their values. For the case of a
single factor of $\Sigma$ the only diagrams up to fourth order in $\Lv_\hy$ are
the ones shown in Figs.~\ref{fig.rate_nSm}(a)--(d). The only fourth-order
contribution from the term with two $\Sigma$ is shown in
Fig.~\ref{fig.rate_nSm}(e). The second insertion of $\Lv_\hy$ counting from the
lower right comes from the upper factor of $\Sigma$ due to the right-most ``Q''
in Fig.~\ref{fig.rate_nSigma}(b).



The diagrams in Figs.~\ref{fig.rate_nSm}(d) and (e) are topologically
equivalent but differ in the time ordering. Diagram (d) has $t\equiv
t_4>t_1>t_2>t_3$, whereas (e) has $t_4>t_1$ and $t_4>t_2>t_3$. In addition, (e)
has an odd number of $\Lv_\hy$ on the lower branch of the upper $\Sigma$,
leading to an additional minus sign from Eq.~(\ref{TCL1.Uexp.4}). If we add (d)
and (e), contributions with $t_1>t_2$ cancel and we obtain  $t_4>t_2>t_1$ and
$t_2>t_3$ and an overall minus sign. Higher-order Diagrams are constructed in
the same manner.

\section{Summary}

In this paper various approaches to the ME for tunneling through
molecules and quantum dots have been discussed and compared. The standard
derivation of the WBR ME relies on two assumptions: (1) Weak
tunneling---allows one to use low-order perturbation theory in the tunneling
amplitude and generally implies the Markov assumption of temporal locality. (2)
The leads form large energy and particle reservoirs---together with the first
assumption and an initial density operator at time $t_0$ of product
form with the leads in equilibrium this allows one to write the
density operator as a product at all times. Assumption (2) must only be
made when calculating expressions of the desired order in $H_\hy$. Making it
globally leads to trivial dynamics.

The Markov approximation does not require one to invoke rapid relaxation due to
electron-electron or electron-phonon interaction in the leads. In fact, under
rather weak conditions, these interactions are shown to be irrelevant for the
tunneling. A short time scale describing the decay of correlation in the leads
emerges naturally, given by the inverse bias or the inverse thermal energy,
whichever is smaller. At low temperatures, this dominates over the contribution
from the finite quasiparticle lifetime. A more careful analysis is needed if
tunneling events are highly correlated in time.

The superoperator derivation of the WBR ME clarifies the
role of the Markov approximation. No assumptions beyond weak tunneling
and an initial density operator of product form are made. The large-reservoir
assumption (2) is not required, beyond this initial condition. The resulting
WBR ME is non-local in time. An explicit Markov approximation is
required to make it local.

The ME of K\"onig \textit{et al.}\cite{KSS96b} is equivalent to the WBR ME to
all orders in tunneling. Its memory kernel, given as a diagrammatic
perturbation series in Ref.~\onlinecite{KSS96b}, has been written down in
superoperator form.

The \textit{T}-matrix approach only gives rates for the \emph{diagonal}
components of the reduced density operator. The large-reservoir assumption of a
product state enters in the guise of statistical independence of dot and lead
states. The \textit{T}-matrix approach gives the same expression for these
rates as WBR theory with the Markov assumption only to leading (second) order
in $H_\hy$, corresponding to Fermi's Golden Rule. This is because in the
\textit{T}-matrix approach one calculates a subtly different quantity than in
the WBR approach.



The TCL ME\cite{ToM75} has been adapted to the transport problem. It is an
exact equation for the dynamic reduced density operator that is local in time
but does not require a Markov assumption. It thus works to arbitrary orders in
perturbation theory. The superoperator $\Sigma(t-t_0)$ playing a pivotal role
in the TCL ME has been given in an explicit form.

A number of technical questions regarding the TCL formalism applied to
transport are still open. First, under what conditions does the inverse of the
superoperator $1-\Sigma(t-t_0)$ exist, which appears in the ME?
Second, if we assume the system to be in a product state with leads in
equilibrium at time $t_0$, can the limit $t_0\to-\infty$ be taken? And third,
is positivity of probabilities satisfied in perturbative approximations to the
TCL ME? Of course, this may be asked for any approach. We
leave these questions for future work.


The assumption of a product state with leads in equilibrium at time $t_0$ is not
required in this approach, but simplifies the results. There are processes of
first order in $H_\hy$ that are physically reasonable but are omitted if we make
this assumption. They can be incorporated by choosing non-standard initial
conditions.

A diagrammatic scheme for generating arbitrary orders in $H_\hy$ in the TCL ME
has been developed. The relation to the diagrams of K\"onig \textit{et
al.}\cite{KSS96b} and thus to the WBR ME is easily seen. Our diagrams have
interesting additional structure, since the projected density operator is
propagated backward in time to make the equation local. The diagrammatic
expansion to fourth order is shown explicitly.

\acknowledgments

The author would like to thank J. Koch, F. Elste, J. K\"onig, and J. P. Ralston
for valuable discussions and helpful comments on the manuscript and the KITP,
Santa Barbara, and the Freie Universit\"at Berlin for their hospitality. This
work was supported in part by NSF Grant No.\ PHY99-07949.


\appendix

\section{The WBR master equation to second order}
\label{app.explicit}

Equation (\ref{GZ.drho.11}) is the WBR ME to second order in
$H_\hy$. We here give two more explicit forms. First, we derive a fully general
expression useful for later comparisons. We
introduce dot states $|m)$ and lead states
$|i\rrangle$ with eigenenergies $E_m$ and $\epsilon_i$, respectively,
open the commutators, and perform the time integral,
\begin{widetext}
\ba
\dot\rho^\dt_{mn} & = & -i\, (E_m-E_n)\, \rho^\dt_{mn} \nonumber \\
& & {}- \pi \sum_{ij} \sum_{pq}
  \Big\{ W_i\, \llangle i|(m| H_\hy |p) |j\rrangle\,
  \llangle j| (p| H_\hy |q) |i\rrangle
   \rho^\dt_{qn} \, \delta(E_p+\epsilon_j-E_q-\epsilon_i)
   \nonumber \\
& & \qquad{}- W_j\, \llangle i| (m|H_\hy|p) |j\rrangle
  \rho^\dt_{pq} \llangle j| (q| H_\hy |n)|i\rrangle
  \, \delta(E_q+\epsilon_j-E_n-\epsilon_i)
  \nonumber \\
& & \qquad{}- W_j\, \llangle i|(m| H_\hy |p) |j\rrangle
   \rho^\dt_{pq} \llangle j| (q|H_\hy|n) |i\rrangle
   \, \delta(E_m+\epsilon_i-E_p-\epsilon_j)
   \nonumber \\
& & \qquad{}+ W_i\, \rho^\dt_{mp} \llangle
  i| (p| H_\hy |q) |j\rrangle
  \llangle j| (q| H_\hy |n) |i\rrangle
  \, \delta(E_m+\epsilon_i-E_q-\epsilon_j)
  \Big\} \nonumber \\
& & + i \sum_{ij} \sum_{pq}
  \bigg\{ W_i\, \llangle i|(m| H_\hy |p) |j\rrangle\,
  \llangle j| (p| H_\hy |q) |i\rrangle
   \rho^\dt_{qn} \, P \frac{1}{E_p+\epsilon_j-E_q-\epsilon_i}
   \nonumber \\
& & \qquad{}- W_j\, \llangle i| (m|H_\hy|p) |j\rrangle
  \rho^\dt_{pq} \llangle j| (q| H_\hy |n)|i\rrangle
  \, P \frac{1}{E_q+\epsilon_j-E_n-\epsilon_i}
  \nonumber \\
& & \qquad{}- W_j\, \llangle i|(m| H_\hy |p) |j\rrangle
   \rho^\dt_{pq} \llangle j| (q|H_\hy|n) |i\rrangle
   \, P \frac{1}{E_m+\epsilon_i-E_p-\epsilon_j}
   \nonumber \\
& & \qquad{}+ W_i\, \rho^\dt_{mp} \llangle
  i| (p| H_\hy |q) |j\rrangle
  \llangle j| (q| H_\hy |n) |i\rrangle
  \, P \frac{1}{E_m+\epsilon_i-E_q-\epsilon_j}
  \bigg\} .
\label{GZ.drho.13}
\ea
\end{widetext}
Here, $W_i\equiv \llangle i|\rho_\ld^0|i\rrangle$ is the probability
to find the leads in state $|i\rrangle$ and $P$ denotes the principal value.

Second, we consider a specific model
with electrons in the leads $\alpha=\mathrm{L},\mathrm{R}$ characterized by
wave vector $\mathbf{k}$ and spin $\sigma$ and with molecular orbitals
enumerated by $\nu$. The hybridization is described by the Hamiltonian
\be
H_\hy = -\frac{1}{\sqrt{N}} \sum_{\alpha\mathbf{k}\sigma\nu} \left(
  t_{\alpha\mathbf{k}\sigma\nu} a^\dagger_{\alpha\mathbf{k}\sigma} c_{\nu\sigma}
  + \mbox{h.c.} \right) ,
\ee
where $N$ is the number of sites in each lead.
$a^\dagger_{\alpha\mathbf{k}\sigma}$ ($c^\dagger_{\nu\sigma}$) creates an
electron in lead $\alpha$ (in molecular orbital $\nu$).
If we insert $H_\hy$ into Eq.~(\ref{GZ.drho.11}), only a single sum
over $\alpha,\mathbf{k},\sigma$ survives, due to the conservation of momentum,
spin, and lead index. We introducing dot states $|m)$ with eigenenergies $E_m$
and open the commutators. We only give the first of eight terms,
the others are analogous:
\ba
\lefteqn{ \dot\rho^\dt_{mn} = -i\, (E_m-E_n)\, \rho^\dt_{mn} 
  - \int_0^\infty \! d\tau\, \tr_\ld \sum_{pq} \frac{1}{N}
  \sum_{\alpha\mathbf{k}\sigma} \sum_{\nu\nu'} } \nonumber \\
& & {}\times \Big\{
  t_{\alpha\mathbf{k}\sigma\nu} a^\dagger_{\alpha\mathbf{k}\sigma}\,
  (m| c_{\nu\sigma} |p)\, e^{-iE_p\tau} e^{-i H_\ld \tau}
  t^\ast_{\alpha\mathbf{k}\sigma\nu'} \nonumber \\
& & \quad{}\times (p| c^\dagger_{\nu'\sigma} |q)\,
  a_{\alpha\mathbf{k}\sigma}\, e^{iE_q\tau} e^{i H_\ld \tau}\,
  \rho^\dt_{qn} \otimes \rho^0_\ld + \ldots \Big\} . \nonumber \\
{}
\ea
The second-order term contains the expression
\be
\tr_\ld a^\dagger_{\alpha\mathbf{k}\sigma} e^{-iH_\ld\tau}
  a_{\alpha\mathbf{k}\sigma} e^{iH_\ld \tau} \rho^0_\ld
  = -i\, G^{<}_{\alpha\mathbf{k}\sigma}(-\tau) .
\ee
All terms contain lesser or greater Green functions,
$G^{<}$ or $G^{>}$, respectively, which describe the lead correlations discussed
in Sec.~\ref{sec.WBR}.
Their Fourier transforms can be expressed in terms of the Fermi function $n_F$
and the spectral function of the leads, $A_{\alpha\mathbf{k}\sigma}(\omega)$, as
\ba
G^{<}_{\alpha\mathbf{k}\sigma}(\omega) & = & i\, n_F(\omega - \mu_\alpha)\,
  A_{\alpha\mathbf{k}\sigma}(\omega - \mu_\alpha) , \\
G^{>}_{\alpha\mathbf{k}\sigma}(\omega) & = & -i\, [1-n_F(\omega - \mu_\alpha)]\,
  A_{\alpha\mathbf{k}\sigma}(\omega - \mu_\alpha) , \qquad
\ea
where $\mu_\alpha$ is the chemical potential of lead $\alpha$. Performing the
integral over $\tau$, we obtain
\ba
\lefteqn{ \dot\rho^\dt_{mn} = -i\, (E_m-E_n)\, \rho^\dt_{mn}
  + i \sum_{pq} \frac{1}{N} \sum_{\alpha\mathbf{k}\sigma} \sum_{\nu\nu'}
  \int \frac{d\omega}{2\pi}} \nonumber \\
& & {}\times A_{\alpha\mathbf{k}\sigma}(\omega-\mu_\alpha)\,
  \bigg\{ \frac{n_F(\omega-\mu_\alpha)}{-\omega+E_p-E_q-i0^+}\,
  t_{\alpha\mathbf{k}\sigma\nu} t^\ast_{\alpha\mathbf{k}\sigma\nu'} \nonumber \\
& & \quad{}\times (m| c_{\nu\sigma} |p)
  (p| c^\dagger_{\nu'\sigma} |q)\, \rho^\dt_{qn} + \ldots
  \bigg\} .
\ea
If we assume that the tunneling amplitudes
$t_{\alpha\mathbf{k}\sigma\nu}\equiv t_{\alpha\sigma\nu}$
do not depend on the wave vector $\mathbf{k}$,
the sum over $\mathbf{k}$ can be performed, noting that the spin-resolved
density of states is given by
$D_{\alpha\sigma}(\omega) = (2\pi V)^{-1} \sum_{\mathbf{k}}
A_{\alpha\mathbf{k}\sigma}(\omega)$.
Here, $V$ is the system volume. This leads to
\begin{widetext}
\ba
\dot\rho^\dt_{mn} & = & -i\, (E_m-E_n)\, \rho^\dt_{mn} \nonumber \\
& & {}- \pi\, \frac{V}{N} \sum_{pq} \sum_{\alpha\sigma\nu\nu'} \Big\{
  D_{\alpha\sigma}(E_p-E_q-\mu_\alpha)\, n_F(E_p-E_q-\mu_\alpha)\,
  t_{\alpha\sigma\nu} t^\ast_{\alpha\sigma\nu'}
  (m| c_{\nu\sigma} |p)(p| c^\dagger_{\nu'\sigma} |q)\, \rho^\dt_{qn}
  + \ldots \Big\} \nonumber \\
& & {}- i\, \frac{V}{N} \sum_{pq} \sum_{\alpha\sigma\nu\nu'} P\! \int d\omega\,
  D_{\alpha\sigma}(\omega-\mu_\alpha)\, \bigg\{
  \frac{n_F(\omega-\mu_\alpha)}{\omega-E_p+E_q}\,
  t_{\alpha\sigma\nu} t^\ast_{\alpha\sigma\nu'}
  (m| c_{\nu\sigma} |p)(p| c^\dagger_{\nu'\sigma} |q)\, \rho^\dt_{qn}
  + \ldots \bigg\} ,
\ea
\end{widetext}
where $P$ indicates a principal value integral.

We find that if the large-reservoir approximation is valid and if the tunneling
amplitudes $t_{\alpha\mathbf{k}\sigma\nu}$ do not depend on $\mathbf{k}$, only
the lead density of states enters.\cite{footnote.notdos} Strong
correlations in the leads, resulting in broad features in the spectral
function, do no affect the tunneling. Under these conditions all information on
$\mathbf{k}$ is lost so that it is only important whether states at a given
energy (and $\alpha$, $\sigma$) exist. As discussed in Sec.~\ref{sub.Markov},
broad features in the spectral function do not invalidate the Markov
approximation made here, since they correspond to rapid processes.

\section{First-order terms}
\label{app.first}

We briefly discuss first-order terms in the ME. Equation
(\ref{TCL1.master2.5}) shows that the only first-order term comes from the
initial condition for $\Q\rho(t_0)$. The origin of the vanishing of first-order
terms is that they contain equilibrium averages of single lead-fermion
operators, cf.\ Eqs.~(\ref{GZ.trHr.2}) and (\ref{TCL1.PHhyP.1}).


The absence of first-order terms might be surprising. Let us consider
$\rho_\dt$ with component $\rho^\dt_{mn}\neq 0$, corresponding to the
superposition of two states with electron number differing by one. If
$(n|H_\hy|m)\neq 0$, a single power of $H_\hy$ is sufficient to lead to a
change in $\rho^\dt_{mm}$ and $\rho^\dt_{nn}$, analogous to the interaction of
a superposition having an oscillating dipole moment with the light field.
Superselection rules\cite{WWW52,Zur82,GKZ95} suggest that superpositions of
states with different charge dephase so rapidly that they are unobservable.
However, this does not help us at a fundamental level, since we would obtain
the same equation if the tunneling fermions were neutral.

A first-order term is present if we expand the equation for the full density
operator, $\rho(t) = e^{-i\Lv(t-t_0)}\, \rho(t_0)$, in powers of $\Lv_\hy$.
Since first-order processes thus exist, while $\P\rho$ does not describe them,
the required information must be contained in $\Q\rho(t)$. If they happen to be
relevant, we must choose $\Q\rho(t_0)\neq 0$.


\begin{thebibliography}{99}
\bibitem{footnote.Born}The term ``Born approximation'' is sometimes used for
the weak-tunneling assumption, specifically for the expansion to second
order\protect\cite{Whi07} or for locality in time\protect\cite{HEM06} (usually
considered part of the Markov approximation), or for the large-reservoir
assumption.\protect\cite{BKP99,BrP02,GoL07} The first usage is more in line
with the Born approximation in scattering theory.
\bibitem{footnote.KSSS}The relation $\Pi(t,t'') = \Pi(t,t')\,\Pi(t',t'')$ is
inconsistent with Eq.~(\protect\ref{KSSS.Dy.4}) since
$\Pi(t,t'')$ contains contributions that are irreducible at time $t'$, whereas
$\Pi(t,t')\,\Pi(t',t'')$ does not. The origin of the problem is that $\Pi$
contains a projection $\P$ or lead trace $\tr_\ld$ at the final time.
\bibitem{footnote.notdos}If $t_{\alpha\mathbf{k}\sigma\nu}$ depends on
$\mathbf{k}$, the density of states is replaced by a weighted integral of the
spectral function over the Brillouin zone, without significantly changing the
conclusions.
\end{thebibliography}
\end{document}